\documentclass[sigplan,twocolumn]{acmart}

\usepackage{tikz}
\usepackage{xcolor}
\newcommand*\circled[1]{\tikz[baseline=(char.base)]{
            \node[shape=circle,fill,inner sep=0.5pt] (char) {\textcolor{white}{#1}};}}
            
\usepackage{amsmath,amsfonts}
\usepackage{algorithm}
\usepackage{algpseudocode}

\usepackage{graphicx}
\usepackage{textcomp}
\usepackage{xcolor}
\usepackage{mdframed}
\usepackage{tabularx}
\usepackage{comment}
\usepackage{booktabs}
\usepackage{bm}
\usepackage{subcaption}
\usepackage{adjustbox}
\usepackage{multirow}
\usepackage{cleveref}

\usepackage{mdframed}
\mdfsetup{
  linecolor=red,  
  linewidth=2pt,  
}

\renewcommand\footnotetextcopyrightpermission[1]{} 
\acmConference[Conference acronym 'XX]{Make sure to enter the correct
  conference title from your rights confirmation emai}{June 03--05,
  2018}{Woodstock, NY}

\begin{document}

\title{GPZ: GPU-Accelerated Lossy Compressor for Particle Data}

\author{Ruoyu Li}
\affiliation{%
  \institution{Florida State University}
  \city{Tallahassee}
  \state{FL}
  \country{USA}}
\email{rl13m@fsu.edu}

\author{Yafan Huang}
\affiliation{%
  \institution{University of Iowa}
  \city{Iowa City}
  \state{IA}
  \country{USA}}
\email{yafan-huang@uiowa.edu}

\author{Longtao Zhang}
\affiliation{%
  \institution{Florida State University}
  \city{Tallahassee}
  \state{FL}
  \country{USA}}
\email{lzhang11@fsu.edu}

\author{Zhuoxun Yang}
\affiliation{%
  \institution{Florida State University}
  \city{Tallahassee}
  \state{FL}
  \country{USA}}
\email{zy24b@fsu.edu}

\author{Sheng Di}
\affiliation{%
  \institution{The University of Chicago}
  \institution{Argonne National Laboratory}
  \city{Lemont}
  \state{IL}
  \country{USA}}
\email{sdi1@anl.gov}

\author{Jiajun Huang}
\affiliation{%
  \institution{University of California, Riverside}
  \city{Riverside}
  \state{CA}
  \country{USA}}
\email{jhuan380@ucr.edu}

\author{Jinyang Liu}
\affiliation{%
  \institution{University of Houston}
  \city{Houston}
  \state{TX}
  \country{USA}}
\email{jliu217@central.uh.edu}

\author{Jiannan Tian}
\affiliation{%
  \institution{Oakland University}
  \city{Rochester}
  \state{MI}
  \country{USA}}
\email{jiannan.tian@uky.edu}

\author{Xin Liang}
\affiliation{%
  \institution{University of Kentucky}
  \city{Lexington}
  \state{KY}
  \country{USA}}
\email{xliang@cs.uky.edu}

\author{Guanpeng Li}
\affiliation{%
  \institution{University of Iowa}
  \city{Iowa City}
  \state{IA}
  \country{USA}}
\email{guanpeng-li@uiowa.edu}

\author{Hanqi Guo}
\affiliation{%
  \institution{The Ohio State University}
  \city{Columbus}
  \state{OH}
  \country{USA}}
\email{guo.2154@osu.edu}

\author{Franck Cappello}
\affiliation{%
  \institution{The University of Chicago}
  \institution{Argonne National Laboratory}
  \city{Lemont}
  \state{IL}
  \country{USA}}
\email{cappello@mcs.anl.gov}

\author{Kai Zhao}
\authornote{Corresponding author}
\affiliation{%
  \institution{Florida State University}
  \city{Tallahassee}
  \state{FL}
  \country{USA}}
\email{kzhao@cs.fsu.edu}



\begin{abstract}
Particle‑based simulations and point‑cloud applications generate massive, irregular datasets that challenge storage, I/O, and real‑time analytics.  
Traditional compression techniques struggle with irregular particle distributions and GPU architectural constraints, often resulting in limited throughput and suboptimal compression ratios. 
In this paper, we present GPZ, a high-performance, error-bounded lossy compressor designed specifically for large-scale particle data on modern GPUs. GPZ employs a novel four-stage parallel pipeline that synergistically balances high compression efficiency with the architectural demands of massively parallel hardware. We introduce a suite of targeted optimizations for computation, memory access, and GPU occupancy that enables GPZ to achieve near-hardware-limit throughput. We conduct an extensive evaluation on three distinct GPU architectures (workstation, data center, and edge) using six large-scale, real-world scientific datasets from five distinct domains. The results demonstrate that GPZ consistently and significantly outperforms five state-of-the-art GPU compressors, delivering up to 8x higher end-to-end throughput while simultaneously achieving superior compression ratios and data quality.
\end{abstract}

\maketitle

\section{Introduction}
\label{sec:intro}

The ever-increasing computational power and advances in sensor precision have driven a surge in particle-data applications, from scientific particle-method simulations to commercial point-cloud technologies. 
Such applications represent a physical or geometrical system as a collection of discrete points, rather than as values on a regular multidimensional grid. 
The flexible structure of discrete points makes it possible to deal with complex, dynamic, and irregular systems, with the cost of massive data volumes.
Modern simulations and sensing technologies can generate datasets containing billions or even trillions of particles or points, making storage, transfer, and real-time analysis increasingly challenging. For example, the New Worlds cosmological simulation running on the Summit supercomputer with Nvidia V100 GPUs~\cite{summit}, produces 70 TB of raw data in a single snapshot~\cite{hacc-new-world}. In another case, the 3D Elevation Program (3DEP) of U.S. Geological Survey~\cite{3dep} has collected topographic data of the U.S. in 3D point cloud format exceeding 200 TBs of storage~\cite{lcp}.

The exponential growth of particle data underscores the urgent need for effective high-performance reduction techniques.
First, Without effective reduction, applications  are largely constrained to downsampling or on-the-fly processing. Downsampling discards as much as 90\% of data for storage, which unavoidably causes significant information loss. On-the-fly processing analyzes data in-situ while sacrificing post-hoc analysis and reproducibility since the data will not be stored.
Second, modern particle datasets may contain billions or trillions of points, such that reduction algorithms must be not only accurate but also extremely fast. If data reduction becomes a computational bottleneck, it will negate the benefits of advanced computing and sensing hardware and introduce unacceptable delays in analysis pipelines.
Third, GPUs are now ubiquitous across HPC clusters and personal computing environments, and many particle-based applications from large-scale simulations on supercomputers to point-cloud rendering on desktops leverage GPU acceleration. Therefore, GPU-optimized reduction technique could preserve end-to-end system performance and fully exploit the capabilities of modern hardware.

Despite these advantages, designing high-performance reduction algorithms for particle data presents several technical challenges. 
First, particle datasets typically exhibit non-uniform distributions and limited spatial and temporal coherence. Such irregularity hinders the effectiveness of most existing compression solutions. Lossless methods typically search for redundancy or repeated patterns—both of which are scarce in particle datasets. Lossy schemes often exploit spatial or temporal correlation, but such correlation is limited in particle data compared to structured meshes.
Second, traditional compression methods optimized for sequential CPU pipelines fail to leverage the massive, fine-grained parallelism and specialized memory hierarchies on GPUs. Simply porting CPU-centric codecs often results in limited speed gains due to poor GPU occupancy, uncoalesced memory accesses, and warp divergence.
Third, there is a fundamental trade-off between compression ratio and throughput on GPUs. Complex decorrelation schemes can enhance compression ratios but incur throughput degradation due to additional data transfers, synchronization barriers, and increased register pressure. Therefore, high-performance designs must carefully balance compression efficiency with the architectural constraints of modern GPU hardware.

In this paper, we propose an effective and high performance GPU-based lossy compression system for particle data. Our contributions are three-fold:
\vspace{-3mm}
\begin{itemize}
    \item We design and implement \textbf{GPZ}, a novel high performance lossy compression system for particle data. Its four-stage parallel pipeline is co-designed to maximize both compression effectiveness on the irregular particle distributions and execution throughput on modern GPU architectures.
    \item We develop a suite of targeted, hardware-aware performance optimizations spanning computation, memory access patterns, and GPU occupancy. These strategies minimize architectural bottlenecks, such as memory latency and register pressure, to achieve state-of-the-art end-to-end throughput.
    \item We conduct an extensive experimental evaluation on diverse, large-scale particle datasets using multiple GPU architectures (workstation, data center, and edge). Our results validate that GPZ robustly outperforms leading GPU compressors in throughput, compression ratio, and reconstruction quality.
\end{itemize}

The remaining of the paper is structured as follows. In \Cref{sec:background}, we discuss the research background and related work. \Cref{sec:objective and overview} define the project objective and the overview of our proposed solution. The solution's algorithmic design is detailed in \Cref{sec:algo design} and the performance optimizations are presented in \Cref{sec:performance optimization}. \Cref{sec:evaluation} presents and discusses the evaluation results. Finally, we draw conclusions in \Cref{sec:conclusion}.

\section{Background and Related Work}
\label{sec:background}

In this section, we discuss two important background, particle data applications and lossy compression, and study the work related to high performance particle compression.

\begin{figure}[ht] \centering
   \centering
    \includegraphics[width=1\linewidth]{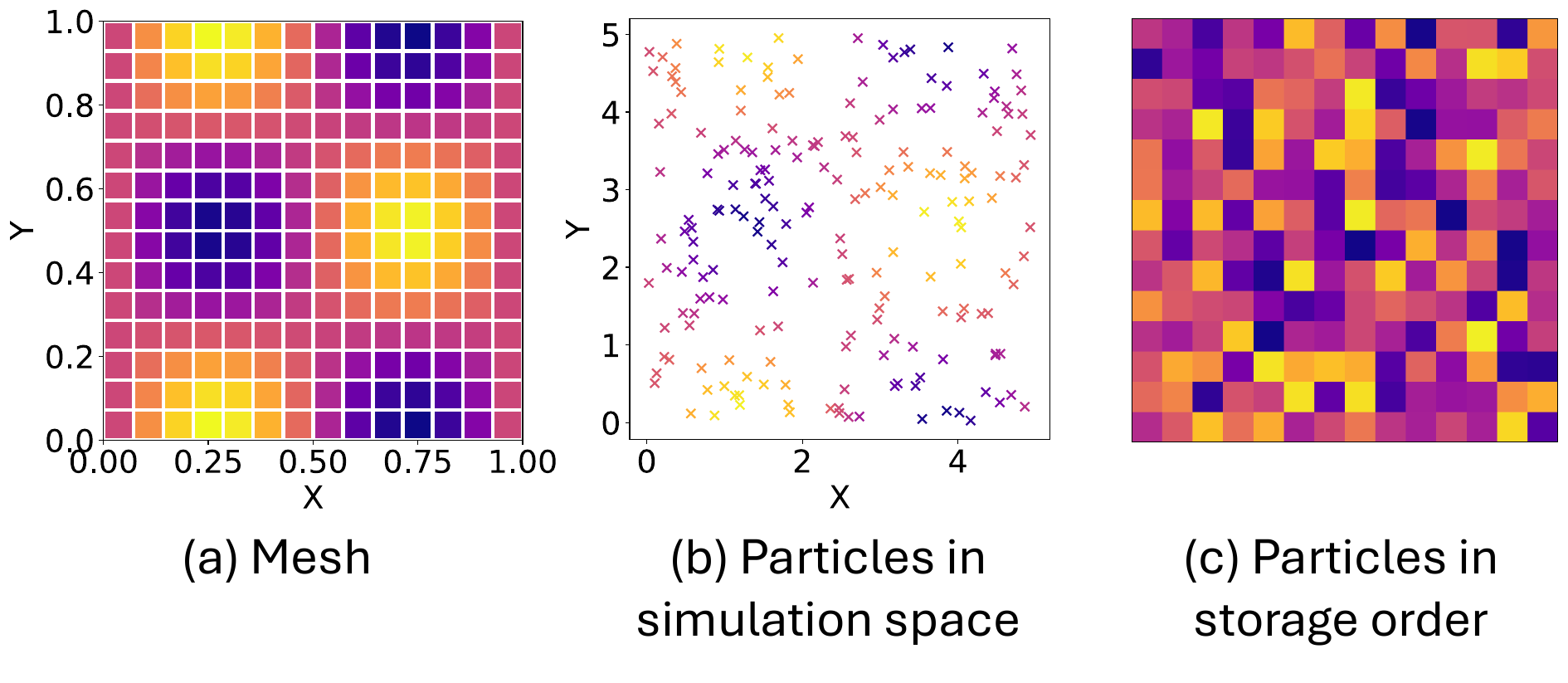}
    \vspace{-8mm}
\caption{Comparison of data coherence in mesh versus particle data. (a) A structured mesh exhibits high coherence. (b) In contrast, particles are irregularly distributed in space. (c) When processing by their storage order, the same particles show low data coherence, which presents a key challenge for compression.}
\label{fig:particle_vs_mesh}
\end{figure}

\subsection{Particle-Data Applications}
\label{sec: background - particles}
Particle-data applications in this paper covers both scientific particle-method simulations and non-scientific point-cloud software, since both domains handle information in particle format which consists of discrete elements.
Compared with mesh (or grid), the particle format provides a flexible and powerful way to model complex or unstructured phenomena. For example, in scientific domain, the LAMMPS molecular dynamics package looks into the microscopic world of atoms and molecules~\cite{lammps} to study material, proteins, or chemical reactions. Such atoms and molecules are represented by particles and simulated with regard to the chemical forces. In non-scientific domain, the LiDAR systems capture the surface of objects with laser and store the information as point clouds for 3D recontruction, mapping, and object recognition. 

\Cref{fig:particle_vs_mesh} visualizes the differences between particle and mesh data formats. \Cref{fig:particle_vs_mesh}(a) is a simple mesh representing a physical property in the 2D space. Each point on this mesh corresponds to a specific scalar value, and these values transition smoothly between neighboring points. In comparison, \Cref{fig:particle_vs_mesh}(b) contains the same number of scalars, but they are organized as discrete particles (or points). By using particles instead of mesh, applications only need to store information in areas of interests, and ignoring background's value. As a result, particle-data applications could handle more complex and larger physical or geometrical systems. This also explains why \Cref{fig:particle_vs_mesh}(b) has a larger 2D space than \Cref{fig:particle_vs_mesh}(a). However, such a discrete and flexible organization dramatically reduces the correlation of neighboring points, as shown in \Cref{fig:particle_vs_mesh}(c) which visualizes particles' values one by one following their storage order. The inconsistency of the storage order and the order in space further decreases such correlation.

\subsection{Error-Bounded Lossy Compression}
\label{sec: background - lossy compression}
Error-Bounded Lossy Compression is a promising solution toward the particle data volume issue, and we discuss the related technologies in this section. 

Lossy compression reduces data volume by discarding noise, insignificant or high‑frequency components and then applying lossless coding (e.g. LZ77, Huffman) to capture the regularities. Purely lossless schemes are already common in particle‑data pipelines. For example, LASzip compresses 3DEP LiDAR terrain scans~\cite{laszip} and LAMMPS offers Zstd output compression. However, they typically achieve limited compression ratios ($\approx 2$) on particle datasets due to discrete data characteristics~\cite{mdz, lcp}. As a result, lossy compression is necessary to push the compression ratio to higher level.

Lossy compression inevitably discards some information from the original data, but error‑bounded methods ensure that the reconstruction error never exceeds a user‑specified tolerance. In this way, they guarantee that any distortion remains negligible for the target application while still maximizing the compression ratio. In what follows, we discuss these techniques in three groups: generic error‑bounded compressors, particle‑data specific methods, and high performance implementations optimized for GPU architectures.

\textbf{Generic error‑bounded compressors} apply domain independent decorrelation techniques to prepare data for lossless encoding, but they often struggle to achieve high compression ratios on irregular particle datasets.
One generic design is to apply statistical models, including linear regression~\cite{sz17}, spline interpolation~\cite{interp}, and neural networks~\cite{coordnet} to predict the value of the variable based on the value of the coordinates. The statistical model will be stored to reconstruct the value such that the storage of original data could be eliminated. To bound the error, the difference between prediction and real value is quantized and stored together with the model. 
Another popular design is to transform data to another domain that is more compressible, and then keep partial of the transformed data based on the error bound. Examples of transformation methods include wavelet transform~\cite{sperr, mgardx} and orthogonal discrete transform~\cite{zfp}. 

\textbf{Particle data specific methods} of error-bounded lossy compression has been developed to exploit the unique structure of particles for higher compression ratios, but most of them do not have performance oriented design and thus suffer from low throughput on modern hardware. 
For point cloud, Google’s Draco~\cite{draco} uses octree spatial partitioning and coordinate quantization for effective compression. The MPEG‐PCC reference codecs TMC13~\cite{tmc13} and TMC2~\cite{tmc2} employ static octree‐plus‐surface‐approximation and temporal residual coding, respectively. 
For molecular dynamics, XTC~\cite{xtc} targets biology systems by exploiting water molecule's local structure for decorrelation, while MDZ~\cite{mdz} improves prediction by leveraging the spatial patterns found in solid materials.
In addition, LCP~\cite{lcp} employs a blockwise spatial decomposition that has proven effective for both scientific simulations and point cloud workflows.

\textbf{High‑performance methods} target GPU architectures with highly parallel, low overhead kernels, but they often produce suboptimal compression ratios (as shown in \Cref{sec:evaluation}) when applied to the discrete particle datasets. 
cuSZp~\cite{cuszp, cuszp2}, FZ-GPU, and PFPL~\cite{pfpl} are three leading solutions in this category. They not only employ lightweight algorithms, such as delta coding (or 1D lorenzo), bit shuffling, and run-length coding, but also incorporate parallel performance optimizations to maximize throughput. For example, cuSZp's decoupled lookback prefix-sum algorithm boosts the speed of assembling the output from GPU thread blocks by eliminating certain dependencies. 
In contrast, cuSZ and cuSZi add more sophisticated steps, such as huffman coding and cubic interpolation prediction, to increase the compression ratios. However such steps either require CPU-side processing or cannot be fully parallelized, which severely degrade the performance in practice (also shown in \Cref{sec:evaluation}).


\section{Objective and Design Overview}
\label{sec:objective and overview}
We start this section by defining the project objectives, followed by an overview of our proposed solution.

\subsection{Project Objectives}
\label{sec:objective}
The objective of this work is to develop an high-performance particle compressor that meets the following three key goals.
We focus on the particle position (coordinate) field, which is consistent with prior particle‑data compressors~\cite{xtc, mdz, lcp}.

\underline{\textit{(i) High compression ratio.}}  We aim to produce higher compression ratios than all other high-performance SOTAs.
Compression ratio is defined as the ratio between the original data size and the compressed data size, and it quantifies the space-saving effectiveness of a compressor. 


\underline{\textit{(ii) High performance.}} We target the highest \textbf{end-to-end} performance among all SOTAs. End-to-end performance covers the execution of the entire compression or decompression process, and is measured by throughput (usually in GB/s).
Many GPU studies, however, report only the raw GPU‐kernel throughput and omit overheads from CPU‐side kernels and internal memory allocations. As a result, their real world throughput can be as low as 2\% of the advertised value~\cite{cuszp2}, which severely limits the applicability in practice.

\underline{\textit{(iii) High compression quality.}} Our solution should not only strictly follow the user‑specified error bounds, but also produce data in the highest quality among all high-performance SOTAs.
Lossy compressors must preserve sufficient quality (or fidelity) so that reconstruction errors do not impair downstream applications. We adopt the widely used rate–distortion analysis which plots Peak Signal‑to‑Noise Ratio (PSNR) against bitrate to quantify the trade‑off between data fidelity and storage cost. Since each dataset contains multiple fields, we first compute the normalized root-mean-square error (NRMSE) for each lossy field, then derive the aggregate PSNR as $-20 \log_{10}(\sqrt{ \frac{1}{n} \sum_{i=1}^{n} \text{NRMSE}_i^2 })$.

In comparison, existing high performance methods are built for regular mesh and yield suboptimal ratios on particle datasets, while existing particle-specific algorithms rely on complex decorrelation that hinders the throughput. Our solution will close this gap by combining particle-data compression with a design fully optimized for parallel execution.

\subsection{Overview of GPZ} 
\label{sec:overview}

\begin{figure}[ht]
\vspace{-2mm}
    \centering
    \includegraphics[width=1\linewidth]{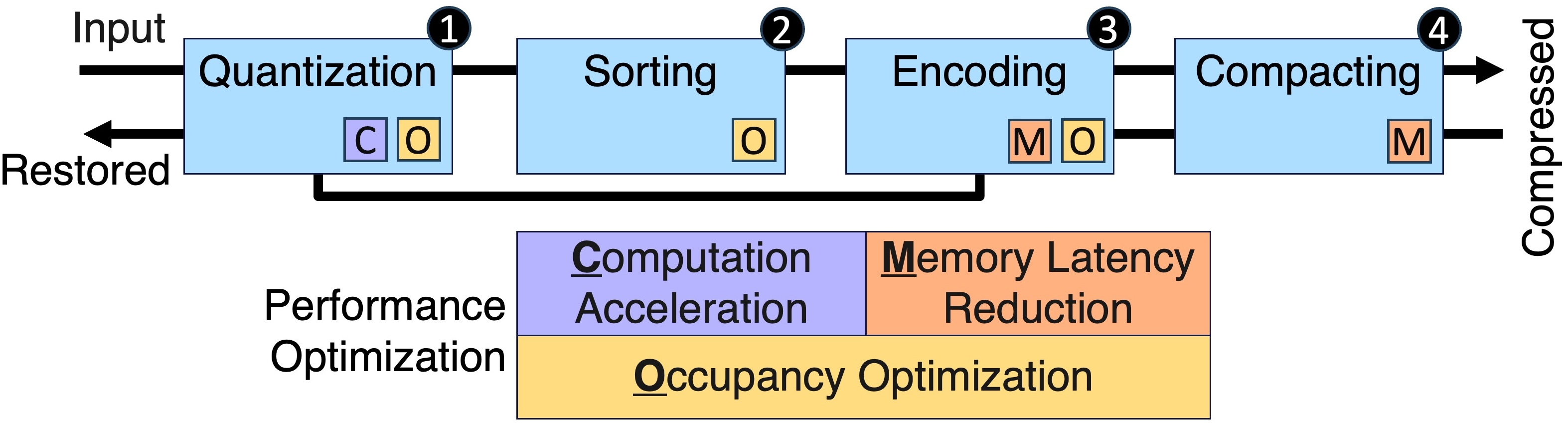}
    \vspace{-2mm}
    \caption{Design overview of our solution}
    \label{fig:module_design}
    \vspace{-2mm}
\end{figure}

We present the high-level overview of our solution GPZ in \Cref{fig:module_design}. 
GPZ consists of a four-stage processing pipeline, with each stage fundamentally designed for parallel execution, and further incorporating tailored performance optimization strategies. 

\begin{figure*}[ht]
    \centering
    \includegraphics[width=0.8\textwidth]{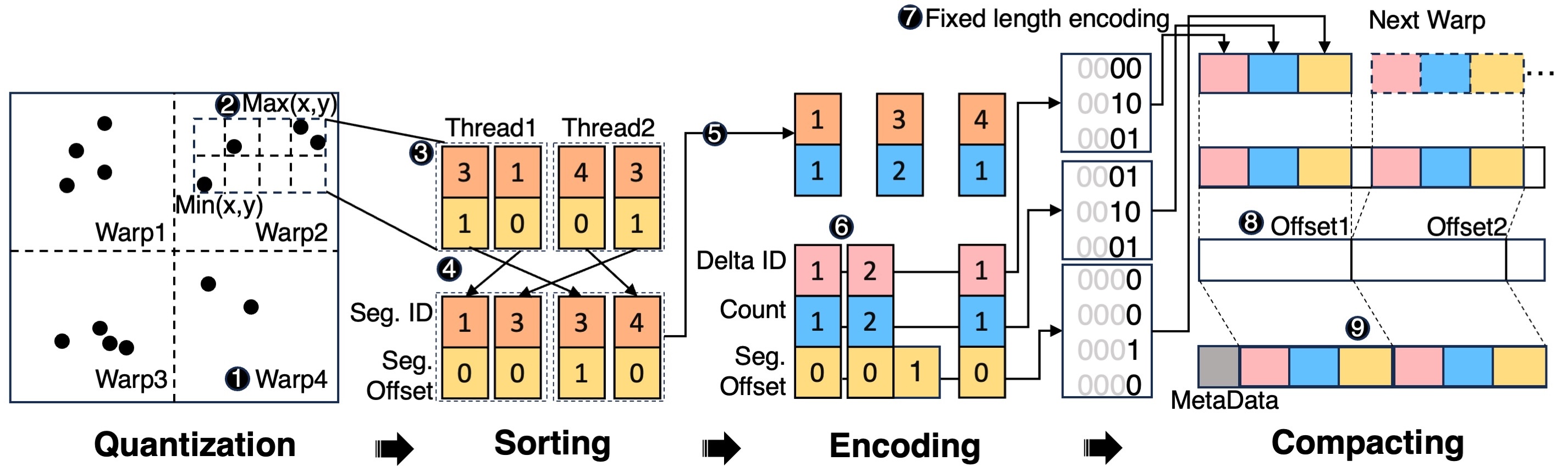}
    \vspace{-2mm}
    \caption{A running example of GPZ compression workflow}
    \label{fig:compression_example}
    \vspace{-2mm}
\end{figure*}

Stages 1–3 process particles in fixed‑size blocks (each block corresponding to a CUDA thread block), while Stage 4 assembles the intermediate outputs from each block into the final compressed format.

The first stage is \textbf{Spatial Quantization} \circled{1} (detailed in \Cref{sec: design - quantization}) which converts each particle's floating-point positions to integers based on its spatial partition while strictly respecting to the user-defined error-bound. 
To achieve this, we partition the spatial domain occupied by the particles into fixed‑size segments. Each particle’s position is then replaced by two integers, its segment ID and offset.
The performance optimization in this stage include selecting the segment size for better GPU occupancy and utilization (see \Cref{sec:performance - occpancy}), and adjusting the floating point math operations for faster computation speed (see \Cref{sec:performance - computation}).

The second stage is \textbf{Spatial Sorting} \circled{2} (detailed in \Cref{sec: design - sorting}) which sorts the particles in each block based on their segment ID. 
As explained in \Cref{sec: background - particles}, particles are not necessarily stored inline with their spatial order, so sorting based on segment will group together particles in the same spatial region and make the subsequent coding step more effective.
To balance compression efficiency and runtime throughput, we perform sorting in each thread block instead of over the whole dataset (see \Cref{sec: design - sorting}), and optimize the amount of hardware resources allocated (see \Cref{sec:performance - occpancy}).


The third stage is \textbf{Encoding} \circled{3} (detailed in \Cref{sec: design - encoding}) which further reduces the data size losslessly.
More specifically, run-length coding and delta coding will be applied to segment IDs to remove redundancy since multiple particles may have same segment ID since they are in the same spatial region, and then bit-plane coding is applied to all data to eliminate the leading zero bits.
In terms of performance optimization, we propose a low-latency memory access strategy (see \Cref{sec:performance - memory}) which significantly improves the throughput in this stage.


The final stage is \textbf{Compacting} \circled{4} (detailed in \Cref{sec: design - compacting}) which assembles the compressed data. 
In particular, all the previous stages handle data independently in block, and this stage collect such data of varying length and put them into a consecutive memory space as the compressed format.
To optimization the performance, we design a three-stage writing strategy (see \Cref{sec:performance - memory}) employing vectorized memory accesses to effectively reduce memory latency.

Decompression is the reverse of the compression steps (without sorting). It first extracts and decodes the bitstream, and then reconstructs the floating‑point values within the specified error bound as decompressed output.

\section{Algorithmic Design}
\label{sec:algo design}

This section presents the algorithmic design of our solution, organized into the four stages outlined in \Cref{sec:overview}. The core innovation lies in the performance‑centric architecture of each stage. 

Our solution divides particles into fixed‑size blocks so that each block can be processed independently and in parallel. 
We target NVIDIA GPUs and implement each particle block as a single CUDA thread block containing exactly one warp (32 threads). This one‑warp‑per‑block mapping ensures high GPU occupancy and throughput by exploiting SIMT execution and warp‑shuffle intrinsics for direct register‑level communication. 
Since other GPU vendors including AMD and Intel adopt similar programming paradigms, our solution can be seamlessly ported to those platforms.

\subsection{Quantization}
\label{sec: design - quantization}
In this stage we partition particles into spatial segments, and replace their floating point coordinates by integers while respecting to the error-bound.

The entire dataset is first spited into blocks and handled in parallel (one block per CUDA warp). Each block contains particles stored nearby, and covers only a subset of the global spatial domain. Within each group, we further partition its occupied region into fixed‑size segments (see \Cref{fig:compression_example} \circled{1}, which limits the range of segment IDs and benefits subsequent lossless coding.

To perform the partition in each block, we need to determine the spatial boundaries by finding the minimum and maximum position values from the particles.  We perform this reduction in two stages for maximum performance -- first within each thread, then across threads in a warp/block.

In particular, we first find the extrema in each thread, then execute the tree-based min-max reduction in each warp, with the help of warp shuffling primitives (\texttt{\_\_shfl\_down\_sync} in CUDA) for direct register-level data exchange between threads in the same warp. 
The resulting \textbf{block-level maxima and minima}, as shown in Figure \ref{fig:compression_example} \circled{2}, are stored in the compressed format and broadcast to all threads.


Next, we convert each particle’s floating‑point coordinates into a segment ID and an integer offset (see \Cref{fig:compression_example} \circled{3}). Suppose particle \(p\) has coordinates \((p_x, p_y, p_z)\). Its segment ID and offset along the \(x\)‑axis are computed as  
\linebreak $ m \cdot \mathit{segId}_x + \mathit{segOffset}_x = (p_x - \mathit{boundary.min}_x )/(2\,\mathrm{eb})$,
\linebreak where \(m\) is the segment size proportional to $\mathit{boundary.max} - \mathit{boundary.min}$. We apply the same computation independently to the \(y\)‑ and \(z\)‑axes, then linearize the three per‑axis IDs into a single 1D segment ID and the three offsets into a single 1D offset for the final encoding.

\subsection{Sorting}
\label{sec: design - sorting}
To exploit spatial coherence and improve the efficiency of subsequent lossless coding, we sort each particle’s (\texttt{segId}, \texttt{segOffset}) pair by segment ID. To maximize throughput, this sort is performed independently within each particle block (each corresponding to a CUDA warp), leveraging low‑latency warp‑level communication on shared-memory.

For example (Figure \ref{fig:compression_example} \circled{4}), 
if Warp 2 initially holds \linebreak \{(3,1),(1,0)\} on Thread 1 and \{(4,0),(3,1)\} on Thread 2, after sorting it holds  \{(1,0),(3,0)\} on Thread 1 and \{(3,1),(4,0)\} on Thread 2.
We implement this using CUB’s \texttt{BlockRadixSort}. By supplying the bit‑width of the maximum segment ID in the block, the sort inspects only the effective bits which reduces the shared‑memory usage.

Sorting more particles per block improves compression ratio but also increases the shared-memory footprint which reduces GPU occupancy and throughput. As detailed in Section \ref{sec:performance - occpancy}, we tuned this trade-off and set each thread to handle 32 points for an optimal balance of compression ratio and processing speed.


\subsection{Encoding}
\label{sec: design - encoding}
We further reduce the size of intermediate data (segment IDs and offsets) with lossless coding strategies, including run-length coding, delta coding, and fix-length coding.

\begin{figure}[ht]
    \centering
    \includegraphics[width=0.7\linewidth]{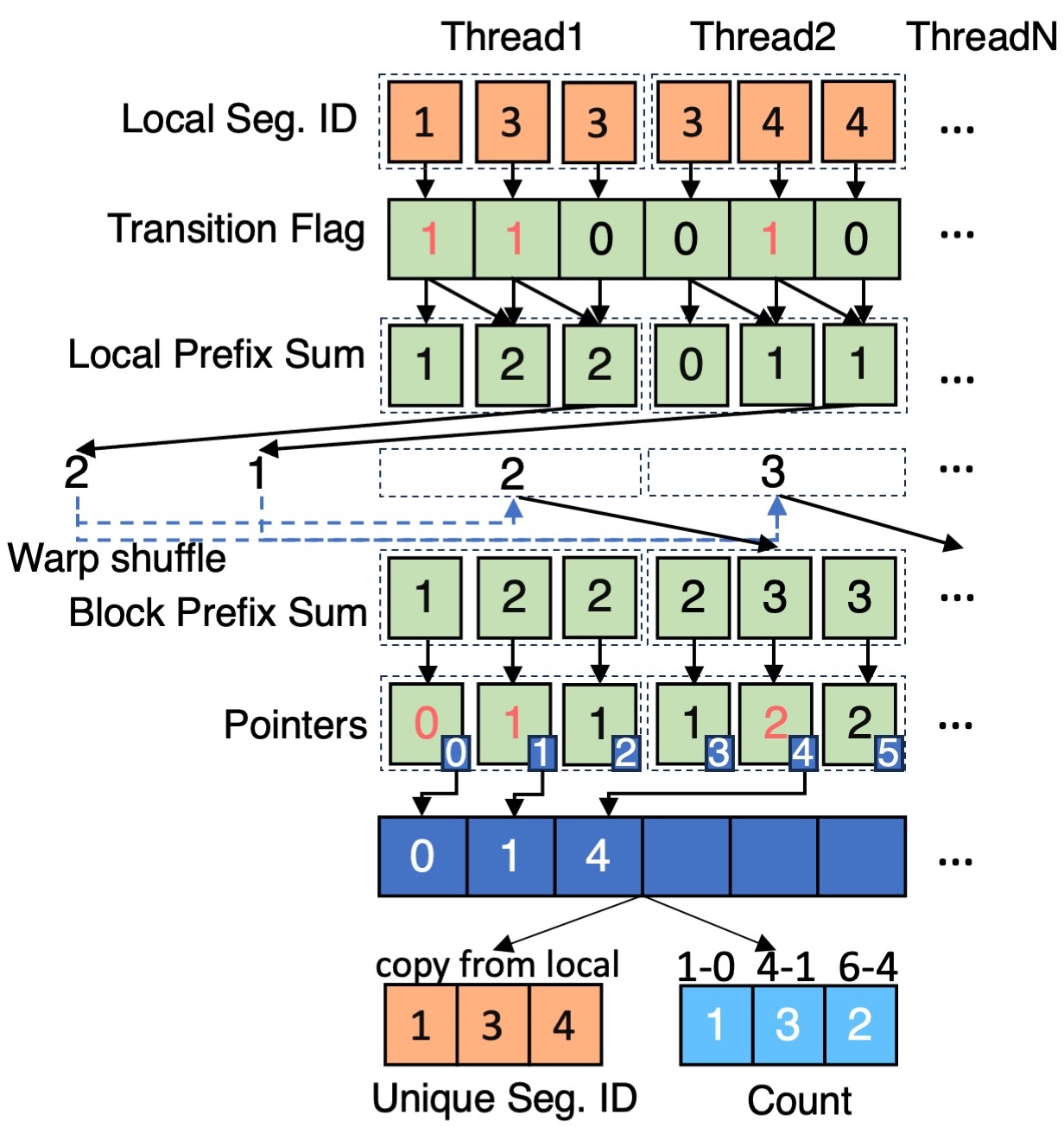}
    \caption{Demonstration of our parallel run-length coding for data holds in multiple threads within a block/wrap}
    \label{fig:encoding_prefixsum}
    \vspace{-4mm}
\end{figure}


We begin by applying parallel run‐length encoding to the sorted segment IDs within each block (as \Cref{fig:compression_example} \circled{5}).  Since sorting produces consecutive, often repeated IDs, run‐length encoding naturally yields a much more compact representation. The primary challenge is to execute this encoding efficiently when each thread holds only a portion of the ID stream.  
To address this, we introduce a novel, low‐overhead, multi‐step algorithm that maximizes parallelism while minimizing  synchronization, as shown in \Cref{fig:encoding_prefixsum}.
In the first step, every thread transmits its last ID to its immediate neighbor using warp‐level primitives, and they independently sets a boolean \textit{Transition Flag} indicating whether the ID changes at its position.
Next, we perform prefix sum in thread level and then in warp/block level to construct the \textit{Pointer} array where each entry in this array points to the slot in the unique‐ID output.
In the final step, each thread independently the indices of its unique IDs (decided by \textit{Transition Flag}) in parallel to a temporal array in shared memory (slot marked by the \textit{Pointer} array), then the final outputs -- the unique IDs and their counts are derived in parallel from the temporal array.

Once the run‐length coding is done, we perform delta encoding on the unique segment IDs (see Figure \ref{fig:compression_example} \circled{6}) by recording the differences between adjacent IDs. Since the IDs are sorted, most deltas are small, thus delta coding greatly reduces data redundancy by exploiting the spatial locality of particles.

In the end, as shown in Figure \ref{fig:compression_example} \circled{7}, we encode the three arrays -- unique segment IDs (in differences), its count, and segment offsets, by fixed length coding which removes redundant leading‑zero bits by representing each value with a uniform bit‑width.

\subsection{Compacting}
\label{sec: design - compacting}

Since particles are handled independently during previous stages, we must consolidate the variable-length data from each block into a continuous memory space as the final output (illustrated in Figure \ref{fig:compression_example} \circled{8}\circled{9}). 

\begin{table}[ht]
\centering
\resizebox{\linewidth}{!}{
\begin{tabular}{|l|c|c|c|c|}
\hline
\multirow{2}{*}{\textbf{Kernel}} & \textbf{Time} & \multicolumn{3}{c|}{\textbf{Throughput}} \\ 
\cline{3-5} 
& \textbf{(ms)} & \textbf{CMP (\%)} & \textbf{MEM (\%)} & \textbf{MEM (GB/s)}  \\
\hline
Q. S. E.     & \multirow{2}{*}{5.27}     & \multirow{2}{*}{43.43} & \multirow{2}{*}{43.41} & \multirow{2}{*}{314.7} \\
with SOTA C. & & & & \\ \hline
Q. S. E.         & 2.59    & 75.4 & 75.4 & 729.9 \\
C. (ours)      & 0.36     & 11.8 & 80.3 & 808.6 \\
\hline
Decompression      & 2.56    & 74.3 & 74.3 & 740.5 \\
\hline
\end{tabular}
}
\caption{Kernel-wise throughput and time breakdown of different compacting approaches. (Q. S. E. C. represent Quantization, Sorting, Encoding, and Compacting, respectively; SOTA C. adopts in-kernel decoupled loopback compacting mechanism;
CMP(\%) and MEM (\%) quantifies GPU resources utilization rate against their theoretical peak sustained rate).}
\label{tab:tp_breakdown}
\vspace{-4mm}
\end{table}


The primary challenge in this consolidation stage is maximizing throughput given the inherent dependency: the final destination of each block's data in continuous memory must be computed before initiating the data copy operation. 
The current SOTA solution \cite{cuszp2} uses an in-kernel inter-block synchronization mechanism (prefix-sum with decoupled look-back) to partially mitigate waiting times for thread-level offset computation. However, such method  frequently uses \texttt{\_\_threadfence()} for inter-block synchronization. Such synchronization causes constant memory stalls and thus significantly impacts the efficiency of our compression pipeline.

To overcome these bottlenecks, we propose a three step compacting strategy. Initially, each warp/block temporarily stores its compressed output from shared memory to global memory. Next, a dedicated device-level prefix-sum kernel computes the global offsets independently. Finally, we launch a third kernel that efficiently copies the data to their respective final positions based on the computed offsets. This approach eliminates inter-block synchronization from the critical execution path, significantly reducing synchronization overhead.
Moreover, thanks to the high compression ratio, the compressed data in each block is considerably smaller than the input. This factor mitigates the overhead from the additional global memory operations. 

As validated by the profiling results presented in Table \ref{tab:tp_breakdown}, our compacting phase is significantly faster than the current SOTA solution~\cite{cuszp2} and occupies only approximately 12\% of the total kernel execution time. Moreover, the measured memory throughput during compaction reaches 809 GB/s on an RTX 4090 GPU which is near the device’s theoretical peak global memory bandwidth.

\section{Performance-Oriented Optimizations}
\label{sec:performance optimization}
This section details our optimizations targeting computation, memory throughput, and hardware-level parallelism (occupancy). We quantify the contribution of each optimization by ablation study in~\Cref{sec: exp - ablation}.

\subsection{Computation Throughput}
\label{sec:performance - computation}

We apply three key optimizations to maximize computational throughput: precision tuning for floating‐point arithmetic, replacement of costly division/modulo operations, and explicit use of fused multiply‐add (FMA) intrinsics.

\underline{\textit{Floating‐point precision.}} We exploit the fact that modern GPUs deliver dramatically higher throughput for 32‑bit arithmetic (e.g., FP64 performance is only 1/64th that of FP32 for NVIDIA RTX 4090). 
Specifically, we retain FP32 if the input dataset is already in FP32 format (the majority of cases; see \Cref{tab:dataset}), and we down‐cast FP64 inputs to FP32 when it is sufficient for the  user‐defined error bound.
Beyond raw ALU throughput gains, this precision reduction also lowers register pressure and memory‐bandwidth usage.

\underline{\textit{Division and Modulo Elimination.}} Division and modulo operators on GPUs incur latencies an order of magnitude (or more) higher than multiplication. To avoid these costs, when computing the segment ID and offset, we refactor constant‐divisor divisions into multiplications by precomputed reciprocals, and replace power‐of‐two modulus operations with bitwise masks. For example, as shown in Table \ref{tab:modular_sass}, in CUDA, the modulo operation is typically implemented using floating-point division and subtraction, which makes it costly in terms of instruction count and register usage. By using a bitwise approach instead, we reduce the instruction count by approximately 80\%, significantly accelerating the computation of segment offsets for each point.

\begin{table}[ht]
\vspace{-3mm}
\centering
\resizebox{\linewidth}{!}{%
\begin{tabular}{|>{\raggedright\arraybackslash}p{3.8cm}|>{\raggedright\arraybackslash}p{8cm}|>{\raggedright\arraybackslash}p{3cm}|}
\hline
\textbf{Implementation} & 
\textbf{Explanation} & 
\textbf{Cost} \\
\hline

\texttt{uint a \% uint b} & 
No native modulo instruction in CUDA; the compiler decomposes the operation into multiple instructions involving floating-point conversion, reciprocal estimation, and reconstruction to simulate integer modulus. 
& $\sim$20+ instructions \newline 5+ registers \\
\hline

\texttt{uint a \& uint b - 1} & 
When \texttt{b} is a power of two, \texttt{a \% b} can be rewritten as a bitwise AND with \texttt{b-1}, enabling the compiler to generate a compact sequence of logic instructions.
& 3--4 instructions \newline 1--2 registers \\
\hline
\end{tabular}
}
\caption{Comparison of modulo implementations cost}
\label{tab:modular_sass}
\vspace{-6mm}
\end{table}

\underline{\textit{FMA operations.}} We explicitly employ the \texttt{\_\_fmaf\_rn} intrinsic to enforce fused multiply-add (FMA) instructions, since we notice compilers may not always automatically convert the code into FMA. 
FMA operations execute multiply and add in one cycle. They not only have higher execution speed but also yield improved numerical accuracy compared with separate operations by avoiding intermediate rounding. 

\underline{\textit{Compute scheduling.}} In our implementation, we launch 32 threads per thread block (i.e., one warp). As a result, we eliminate most uses of \_\_syncthreads() when assigning values to shared memory, and instead utilize \_\_shfl\_sync to directly broadcast register-held values across threads, thereby avoiding unnecessary synchronization overhead. Additionally, we restructure conditional logic using ternary operators and data-dependent computation to preserve uniform control flow and mitigate warp divergence.


\subsection{Memory Throughput}
\label{sec:performance - memory}
Global memory access often dominates execution time when processing large datasets. Therefore, we pay careful attention to memory access patterns at the implementation level to ensure that reads and writes to global memory are coalesced.

 \underline{\textit{Memory coalescing.}} Memory coalescing refers to the process by which a GPU combines multiple threads' global memory accesses into as few transactions as possible, improving memory bandwidth utilization by aligning and merging adjacent accesses within a warp. For example, during loop-based read and write operations, each thread in a warp accesses a distinct 4-byte word, and the warp-level access pattern ensures aligned and contiguous global memory transactions—thereby maximizing memory throughput. In Table \ref{tab:soa_aos_sass}, a simple benchmark focusing solely on write operations demonstrates that coalesced access can achieve up 1.6× higher memory throughput compared to strided access when writing data to global memory.

\begin{table}[h]
\centering
\resizebox{\linewidth}{!}{%
\begin{tabular}{|>{\raggedright\arraybackslash}p{4cm}|>{\raggedright\arraybackslash}p{8cm}|>{\raggedright\arraybackslash}p{3cm}|}
\hline
\textbf{Implementation} & 
\textbf{Explanation} & 
\textbf{Observed Throughput (GB/s)} \\
\hline

\texttt{output[tid] = data[tid]} & 
Threads in a warp write consecutive 4-byte addresses, allowing memory coalescing and burst transactions.
& 401.75 GB/s \\
\hline

\texttt{output[tid * STRIDE] = data[tid * STRIDE]} & 
Addresses are strided across warp, resulting in uncoalesced memory transactions and increased DRAM traffic.
& 254.85 GB/s \\
\hline
\end{tabular}
}
\caption{Comparison of Coalesced and Strided memory write operation in two simplified CUDA kernels with 1024 * 1024 elements and 256 thread group size on RTX 4090}
\label{tab:soa_aos_sass}
\vspace{-8mm}
\end{table}

\subsection{Occupancy}
\label{sec:performance - occpancy}
We aims to improve hardware-level parallelism by increasing the occupancy of our solution. 
Occupancy, defined as the ratio of active warps per streaming multiprocessor (SM) to the maximum number of warps the SM can support, is a key indicator of how effectively a GPU can hide instruction and memory latency by exploiting hardware‐level parallelism.  
We maximize occupancy by carefully managing three primary factors: shared memory usage,  register allocation, and the degree of loop unrolling.

\underline{\textit{Shared memory usage.}}
We minimize per-block shared memory such that more blocks can reside on the same SM.
In addition, a smaller shared-memory allocation can enlarge the effective L1 cache which reduces data access latency, since modern GPUs often use the same physical resources for shared memory and L1 cache.
In our solution, we measure a per‐block shared‐memory requirement of 7.36 KB when using 32‑bit intermediate values, versus 11.36 KB for 64‑bit values (this includes temporary arrays in the sorting stage and auxiliary buffers in the encoding stage).
As a result, we default to 32-bit intermediates unless 64-bit is necessary for value range or precision. This decision is inline with the floating-point precision optimization in \Cref{sec:performance - computation} while deriving from difference reasons (computation versus occupation).



\underline{\textit{Registers allocation.}}
We prioritize register over shared memory for higher speed, but control the usage to a certain degree to avoid associated penalties. 
Registers are the fastest on-chip storage, and replacing shared-memory accesses with register reads/writes can boost I/O throughput. However, excessive register allocation per thread can lower occupancy or even trigger register spilling to global memory which incurs significant performance penalties.
To manage this trade-off, we constrain register usage via the \texttt{--maxrregcount} compilation flag (setting it to 168 on RTX 4090/L4 and 128 on H100) to ensure that each SM maintains a healthy number of active warps without register spills.


\underline{\textit{Loop unrolling.}}
We also optimize the degree of loop unrolling to improve performance. 
Loop unrolling reduces branch divergence and increases instruction-level parallelism (ILP). However, excessive unrolling leads to increased register usage which may degrade performance due to reduced occupancy or register spilling. 
In our solution, we carefully apply unrolling to selected loops with a small number of iterations (e.g., unrolling loops less than 32 iterations), in order to benefit from fewer branch checks and more ILP while avoiding the negative impact of aggressive unrolling.


\section{Experimental Evaluation}
\label{sec:evaluation}
In this section, we describe the experimental setup and evaluate our solution on six datasets against five state-of-the-art high performance compressors.

\subsection{Experimental Setting}
\subsubsection{Platforms}
Our experiments are conducted on three distinct platforms to evaluate compressor performance across a range of GPU architectures, power profiles, and deployment scenarios, as detailed in Table \ref{tab:platform}. 
The first platform is a high-performance \textbf{workstation} equipped with an Intel Core i9-13900K CPU, 188 GB of RAM, and an NVIDIA GeForce RTX 4090 GPU. The second is a server instance featuring an Intel Xeon Platinum 8468 CPU (16 vCPUs), 200 GB of RAM, and an NVIDIA H100 SXM GPU, designed for \textbf{large-scale data center} computations. The third platform is a node with an AMD TURIN CPU, 128 GB of RAM, and an NVIDIA L4 GPU, tailored for energy-efficient \textbf{edge and inference} workloads. 
This diverse hardware selection ensures a comprehensive assessment of each compressor's viability for different scientific computing environments.

\begin{table}[ht]
    \centering
    \resizebox{\linewidth}{!} {
    \begin{tabular}{cccccc}
    \toprule
    \textbf{GPU} & \multicolumn{2}{c}{\textbf{Memory}} & \textbf{SM} & \multicolumn{2}{c}{\textbf{Frequency}}  \\
    \textbf{Type} & \textbf{Size} & \textbf{Bandwidth} & \textbf{Total} & \textbf{SM} & \textbf{DRAM} \\
    \midrule
    RTX 4090 & 24GB &1008GB/s & 128 & 2520MHz & 10501MHz \\
    H100 SXM & 80GB &3.35TB/s & 132 & 1980MHz & 2619MHz \\
    L4 & 24GB &300GB/s   &58 & 2040MHz & 6251MHz \\
    \bottomrule
    \end{tabular}}
    \caption{GPU hardware specifications}
    \label{tab:platform}
    \vspace{-8mm}
\end{table}

\subsubsection{Datasets}
The evaluation covers six real-world particle datasets from five distinct domains, with details shown in Table \ref{tab:dataset}. Most of the datasets are extremely large (in TB and even PB level) which could not be compressed at once, so we randomly select a portion for evaluation. Note that all compressors are evaluated based on the same selected portion.

\begin{table}[ht]
    \centering
    \resizebox{\linewidth}{!} {
    \begin{tabular}{cccccc}
    \toprule
    \textbf{Dataset} & \textbf{Domain} & \textbf{\# Particles Tested} & \textbf{Total Size}  \\
    \midrule
    USGS (3DEP)~\cite{3dep} & Geology  &734,077,453 & > 200 TB \\
    OuterRim (HACC)~\cite{hacc-outerrim} & Cosmology  &83,953,207 & > 5 PB \\
    NewWorlds (HACC)~\cite{hacc-new-world} & Cosmology  &143,735,721 & > 7 PB\\
    WARPX~\cite{warpx} & Plasma Physics  &273,027,788  & > 8 TB\\
    LAMMPS~\cite{lammps} & Material science &67,108,864 &  & \\
    XGC Poincaré~\cite{ren2024particle} & Plasma Physics & 13,077,313 & \\
    \bottomrule
    \end{tabular}}
    \caption{Examples of particle data in various domains}
    \label{tab:dataset}
    \vspace{-8mm}
\end{table}

\subsubsection{Baselines}

We compare GPZ with five state-of-the-art GPU-based lossy compressors, including cuSZp2, PFPL, FZ-GPU, cuSZ, and cuSZ-i.  

\underline{\textit{cuSZp2}} (git version 4f99d6d) \cite{cuszp2-git}: cuSZp is an ultra-fast GPU error-bounded lossy compressor. It has two modes of encoding (plain mode and outlier mode) that users need to manually select. We choose the outlier mode as it demonstrates better performance in general for particle datasets.

\underline{\textit{PFPL}} (git version d829bf0) \cite{pfpl-git}: PFPL is a extreme-fast error-bounded lossy compressor that can produce bit-for-bit identical compressed streams on CPUs and GPUs.

\underline{\textit{cuSZ}} (git version 6218452) \cite{cusz-git}: cuSZ is the GPU version of the SZ2~\cite{sz17} algorithm, aiming to improve SZ's throughput on heterogeneous HPC systems. It uses the lorenzo method~\cite{sz17} for decorrelation and Huffman for coding.

\underline{\textit{FZ-GPU}} (git version e8ec990) \cite{fzgpu-git}: FZ-GPU is a high speed lossy compressor inspired by cuSZ. It reaches higher speed than cuSZ with kernel fusion and replacing Huffman coding by bit shuffle with zero elimination.

\underline{\textit{cuSZ-i}} (git version 68a3b8a)\cite{cuszi-git}: cuSZ-i is a GPU-based error-bounded lossy compressor that leverages interpolation-based prediction~\cite{interp} to significantly improve compression ratio and quality over Lorenzo-based methods.

To our knowledge, no GPU-accelerated lossy compressor exists for large-scale particle data. We therefore select five state-of-the-art compressors designed for generic scientific data as our baselines. However, their general-purpose nature leads to significant robustness issues in our tests. PFPL fails on datasets exceeding 2 GB; cuSZp2 produces verification errors on the H100; FZ-GPU and cuSZ could not handle our largest files; and cuSZ-i lacks 1D data support. These failures highlight a clear need for solutions like GPZ, built to handle the unique scale and format of particle data.




\subsection{Evaluation Results and Analysis}
The evaluation covers four aspects -- ablation study, execution speed, compression ratio, and data quality.
To be more specific, the ablation study confirm the effectiveness of our performance optimizations. 
The speed test confirms GPZ's high execution performance on various platforms.
The compression ratio comparison demonstrates GPZ leads to the smallest footprint under same error-bound, and the quality comparison shows GPZ leads to the best fidelity under same bit rate budget.

\subsubsection{Ablation study}
\label{sec: exp - ablation}
\begin{figure}[ht]
    \centering
    \includegraphics[width=\linewidth]{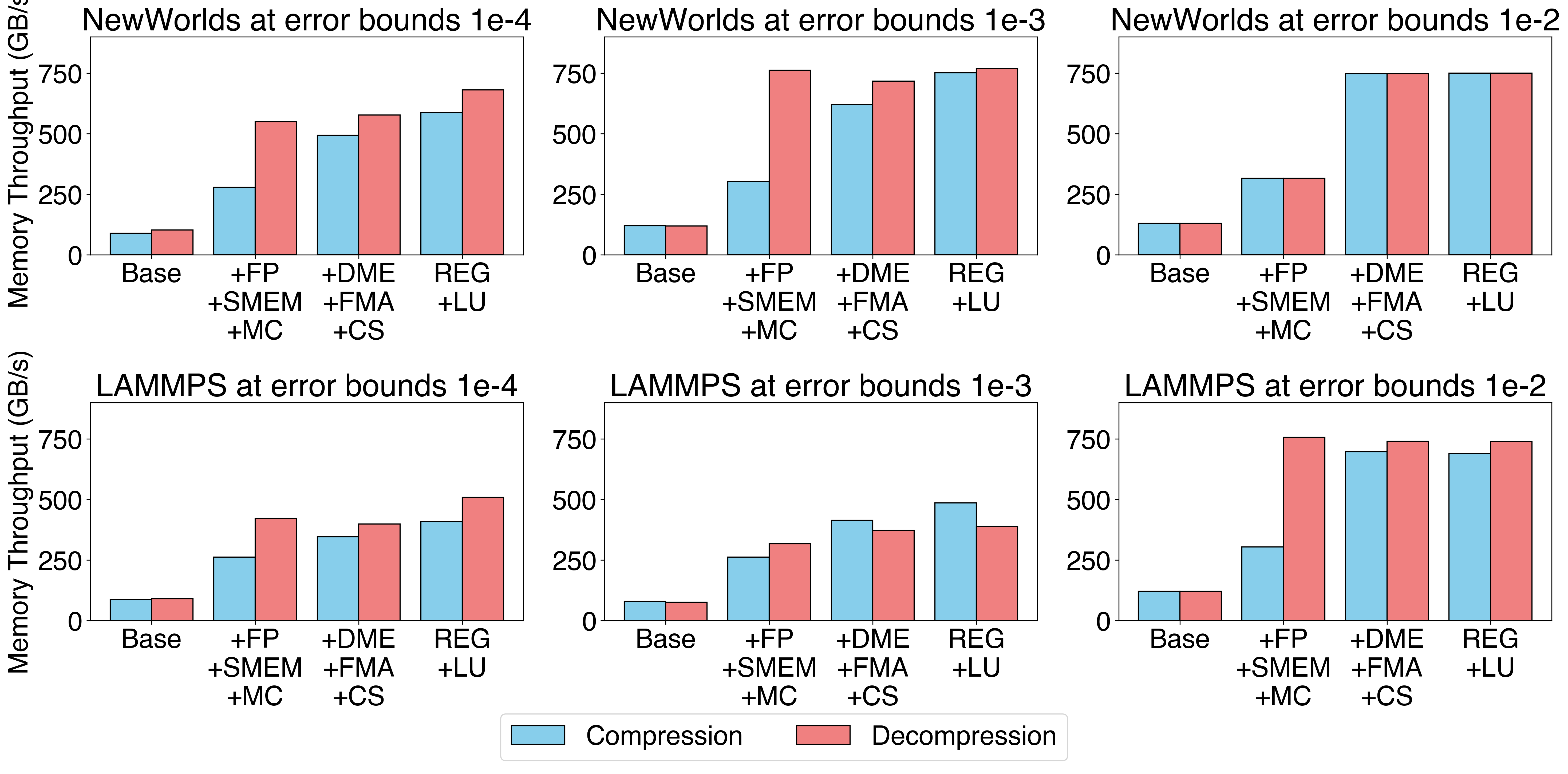}
    \caption{Ablation study of our optimization strategies in~\Cref{sec:performance optimization}. (FP: Floating-point precision; SMEM: Shared memory usage; MC: Memory coalescing; DME: Division and Modulo Elimination;  FMA: FMA operations; CS: Compute scheduling; REG: Registers allocation; LP: Loop unrolling)}
    \label{fig:ablation_study}
\end{figure} 

In Figure \ref{fig:ablation_study}, we present an ablation study to quantify the impact of our key optimizations discussed in~\Cref{sec:performance optimization}. Each group in the figure incrementally adds new techniques to show their cumulative effect. On RTX 4090, these optimizations lead to a dramatic increase in execution speed, improved from 89 GB/s to 588 GB/s in compression and increased from 102 GB/s to 656 GB/s in decompression, 

\subsubsection{Execution Speed}

\begin{figure*}[ht]
    \centering
    \includegraphics[width=\textwidth]{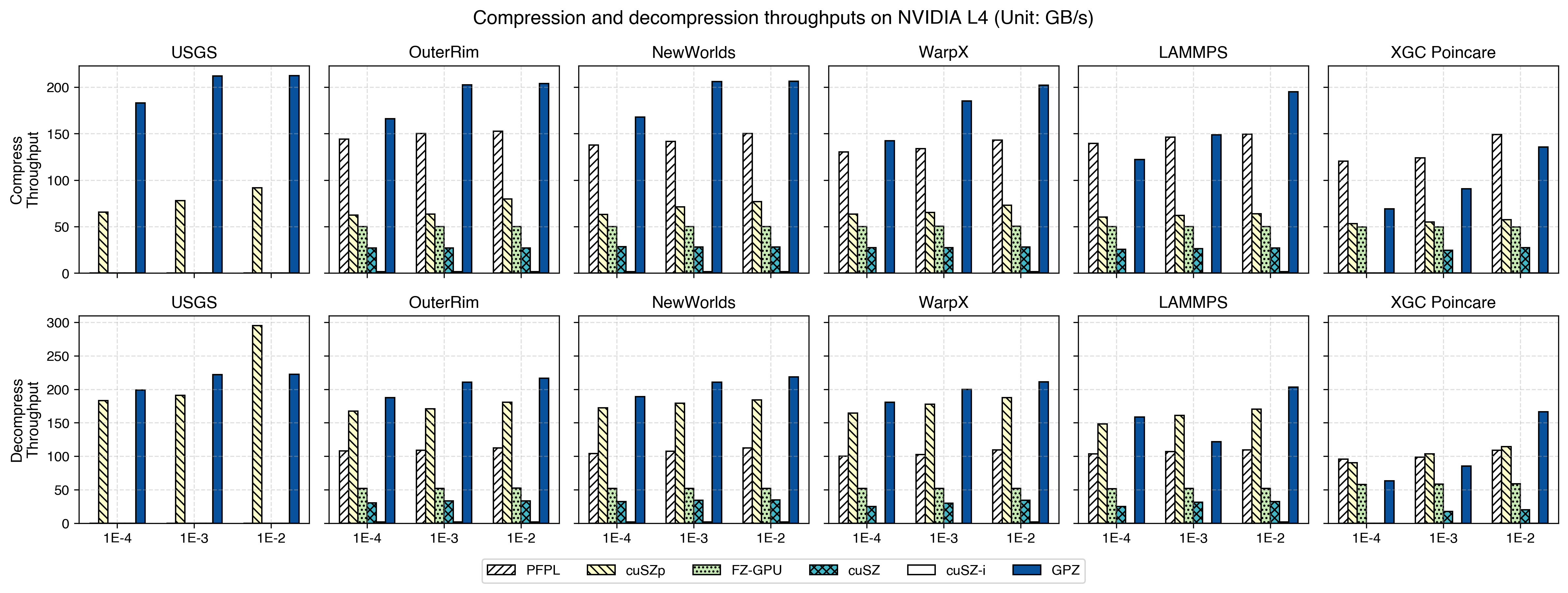}
    \vspace{3mm}
    \includegraphics[width=\textwidth]{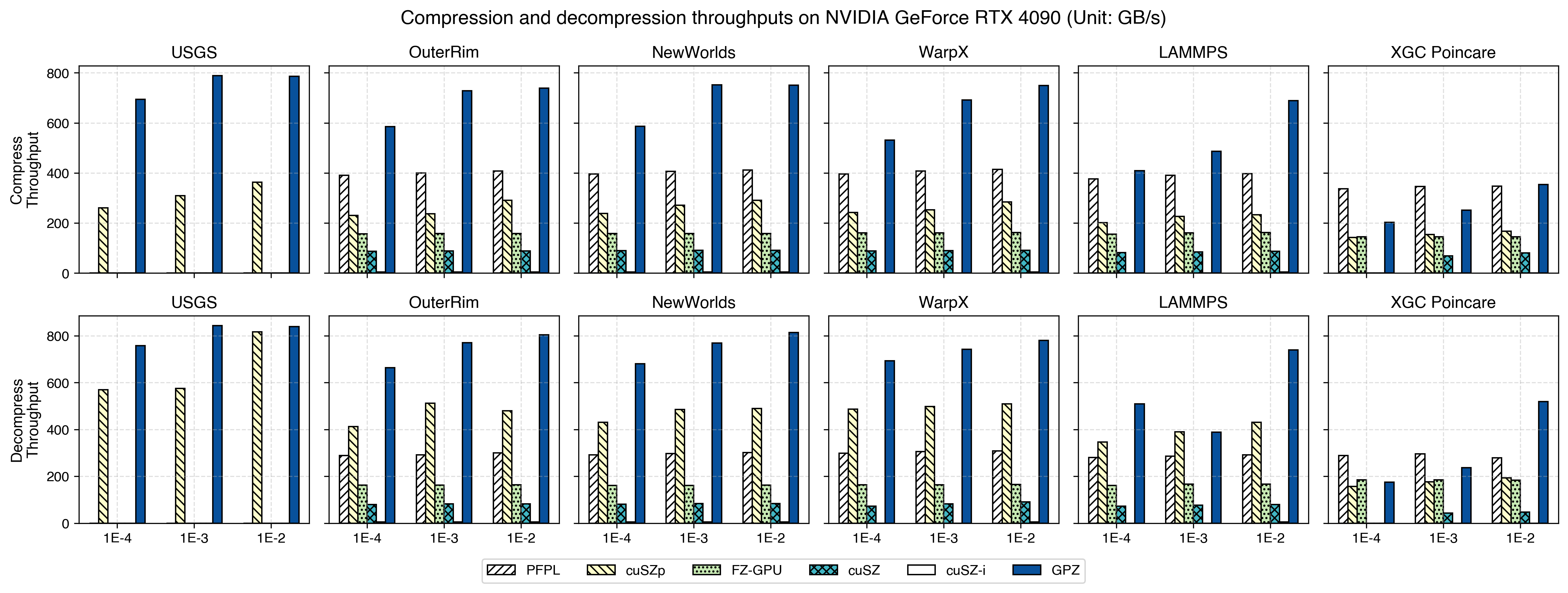}
    \vspace{3mm}
    \includegraphics[width=\textwidth]{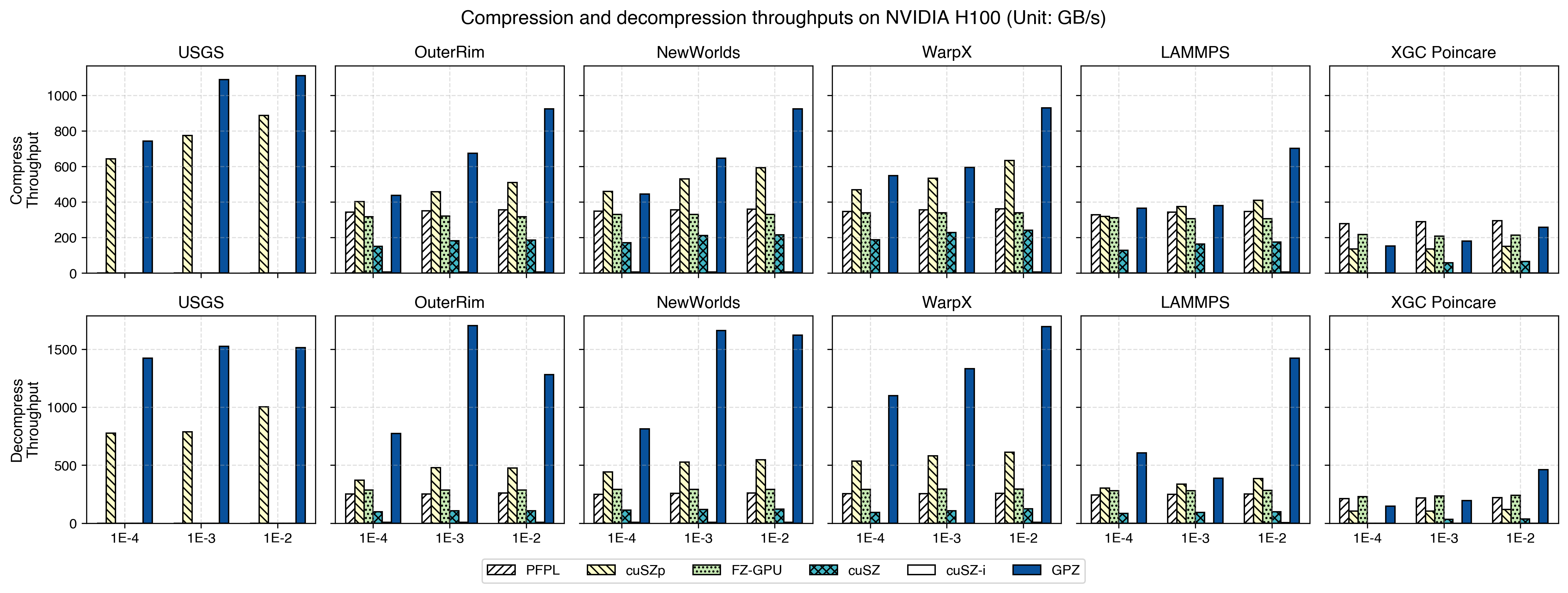}
    \vspace{-3mm}
    \caption{End-to-end throughput on (from top to bottom) NVIDIA L4, RTX 4090, and H100 GPUs. Results are shown for six datasets across four different error bounds. Higher bars indicate better performance. Missing bars denote compressor failures.}
    \label{fig:throughput}
    \vspace{-3mm}
\end{figure*}

We test GPZ on three diverse GPU architectures, including a consumer RTX 4090, a data center H100, and an edge-focused L4, to demonstrate its performance and adaptability across different hardware classes, power consumption, and deployment scenarios. 
We evaluate performance using a comprehensive End-to-End throughput measurement. This includes the entire lifecycle of GPU compression: auxiliary memory allocation, GPU kernel execution, CPU-side processing (if any), and memory deallocation. This approach provides a more realistic measure of practical performance than metrics that only report GPU kernel execution time (as is common in some baseline papers~\cite{cuszi-git, cusz}), since overheads like CPU-side processing maybe significantly higher than GPU kernel time~\cite{cuszp2}.

The results, presented in Figure~\ref{fig:throughput}, demonstrate that GPZ consistently and significantly outperforms all baselines across the three distinct GPU architectures, delivering a 1x to 8x speedup. On average, GPZ achieves compression throughputs of 169 GB/s, 598 GB/s, and 616 GB/s on the L4, RTX 4090, and H100 GPUs, respectively. The corresponding average decompression throughputs are 181 GB/s, 651 GB/s, and 1091 GB/s.

The missing bars in the figures correspond to compressor failures. As noted in our baselines section, these runtime errors were common among the general-purpose compressors when handling our large-scale particle data, with cuSZ-i in particular failing on most test cases. We also note that for the XGC Poincaré dataset, some baselines occasionally achieve higher throughput. However, in these specific cases, GPZ provides a 50\% to 600\% higher compression ratio than the next-best competitor. This superior rate-distortion performance makes GPZ a more practical choice, as it drastically reduces the storage footprint without a significant performance trade-off.

\subsubsection{Compression Ratio}

\begin{table}[ht]
    \centering
    \definecolor{myblue}{rgb}{0.3, 0.5, 0.9}
    \footnotesize 
    \resizebox{\linewidth}{!} {
    \begin{tabular}{c|c|cccccc|c}
    \toprule
    \rotatebox{90}{\textbf{Dataset}} &
    \rotatebox{90}{\textbf{Err. Bound}} & 
    \rotatebox{90}{\textbf{PFPL}} & 
    \rotatebox{90}{\textbf{cuSZp}} & 
    \rotatebox{90}{\textbf{FZ-GPU}} & 
    \rotatebox{90}{\textbf{cuSZ}} & 
    \rotatebox{90}{\textbf{cuSZ-i}} & 
    \rotatebox{90}{\textbf{GPZ}} & 
    \rotatebox{90}{\textbf{Advant.\%}} \\
    \midrule
    
\multirow{3}{*}{\rotatebox{90}{USGS}}
    & 1e-4 & N/A        & 10.63  & N/A   & N/A & N/A   & \textbf{\textcolor{myblue}{15.08}} & 41.86\% \\
    & 1e-3 & N/A        & 18.13  & N/A   & N/A   & N/A   & \textbf{\textcolor{myblue}{217.51}} & 1099.72\% \\
    & 1e-2 & N/A        & 30.53  & N/A   & N/A   & N/A   & \textbf{\textcolor{myblue}{219.42}} & 618.7\% \\[4pt]
              \midrule        
\multirow{3}{*}{\rotatebox{90}{OuterRim}}    
    & 1e-4 & 8.01       & 6.60   & 5.85  & 10.31 & 4.75  & \textbf{\textcolor{myblue}{10.63}} & 3.10\% \\
    & 1e-3 & 26.60      & 13.03  & 10.77 & 25.73 & 6.29  & \textbf{\textcolor{myblue}{131.58}} & 394.66\% \\
    & 1e-2 & 152.36     & 20.10  & 23.70 & 29.86 & 6.37  & \textbf{\textcolor{myblue}{219.41}} & 44.01\% \\[4pt]
    \midrule
\multirow{3}{*}{\rotatebox{90}{NewWorlds}}    
    & 1e-4 & 9.60       & 7.21   & 5.49  & \textbf{\textcolor{myblue}{13.22}} & 5.11  & \underline{10.88} & -17.64\% \\
    & 1e-3 & 31.26      & 12.09  & 8.21  & 26.25 & 6.11  & \textbf{\textcolor{myblue}{97.09}} & 210.59\% \\
    & 1e-2 & 81.30      & 17.22  & 15.48 & 30.13 & 6.35  & \textbf{\textcolor{myblue}{204.79}} & 151.89\% \\[4pt]
    \midrule
\multirow{3}{*}{\rotatebox{90}{WarpX}}    
    & 1e-4 & 7.00  & 5.14   & 4.21  & \textbf{\textcolor{myblue}{8.00}}  & N/A   & \underline{6.15} & -23.12\% \\
    & 1e-3 & 13.20      & 8.41   & 6.58  & 13.02 & N/A   & \textbf{\textcolor{myblue}{21.11}} & 59.92\% \\
    & 1e-2 & 33.69      & 14.00  & 12.50 & 24.03 & 6.14  & \textbf{\textcolor{myblue}{60.71}} & 80.20\% \\[4pt]
    \midrule
\multirow{3}{*}{\rotatebox{90}{LAMMPS}}    
    & 1e-4 & 4.16       & 4.05   & 3.52  & 4.58  & N/A   & \textbf{\textcolor{myblue}{5.17}} & 12.88\% \\
    & 1e-3 & 7.67       & 6.87   & 5.22  & \textbf{\textcolor{myblue}{9.73}}   & N/A   & \underline{8.99} & -7.61\% \\
    & 1e-2 & 22.22      & 13.70  & 9.46  & 24.12 & 6.26  & \textbf{\textcolor{myblue}{33.80}} & 40.14\% \\[4pt]
    \midrule
\multirow{3}{*}{\rotatebox{90}{\shortstack{XGC\\Poincaré}}}
    & 1e-4 & 2.38       & 2.41   & 2.52  & N/A   & N/A   & \textbf{\textcolor{myblue}{3.81}} & 51.19\% \\
    & 1e-3 & 3.20       & 3.21   & 3.42  & 3.52   & N/A   & \textbf{\textcolor{myblue}{7.68}} & 118.18\% \\
    & 1e-2 & 4.90       & 4.73   & 5.13  & 5.53  & N/A   & \textbf{\textcolor{myblue}{42.32}} & 665.28\% \\
    \bottomrule
    \end{tabular}
    }
    \caption{Compression ratio comparison across different datasets and error bounds (last column is the advantage of GPZ over the second best)}
    \label{fig:compression_ratio}
\end{table}

The compression ratio (CR) is a key evaluation metric, as the fundamental goal of compression is to reduce data size and thus ease the demands on storage, transfer, and processing. As demonstrated in~\Cref{fig:compression_ratio}, GPZ consistently achieves higher compression ratios than all other SOTA compressors, thanks to the effective algorithm we present in~\Cref{sec:algo design}.
For the three limited cases when GPZ slightly underperforms cuSZ (7\% - 23\% lower compression ratio), it substantially outperforms cuSZ in execution speed (with a average of 3x to 6x gain on various platforms), as a result, GPZ is still the best option in those cases.
In addition, most compressors, except for cuSZp and GPZ, fail on the USGS dataset due to its large size, which yields the importance of implementing compressors for large-scale datasets.

\subsubsection{Data Quality}
We assess data quality from two perspectives: quantitative rate-distortion analysis and qualitative visual inspection of the decompressed data.

\begin{figure}[ht]
\vspace{-4mm}
    \centering
    \includegraphics[width=0.8\linewidth]{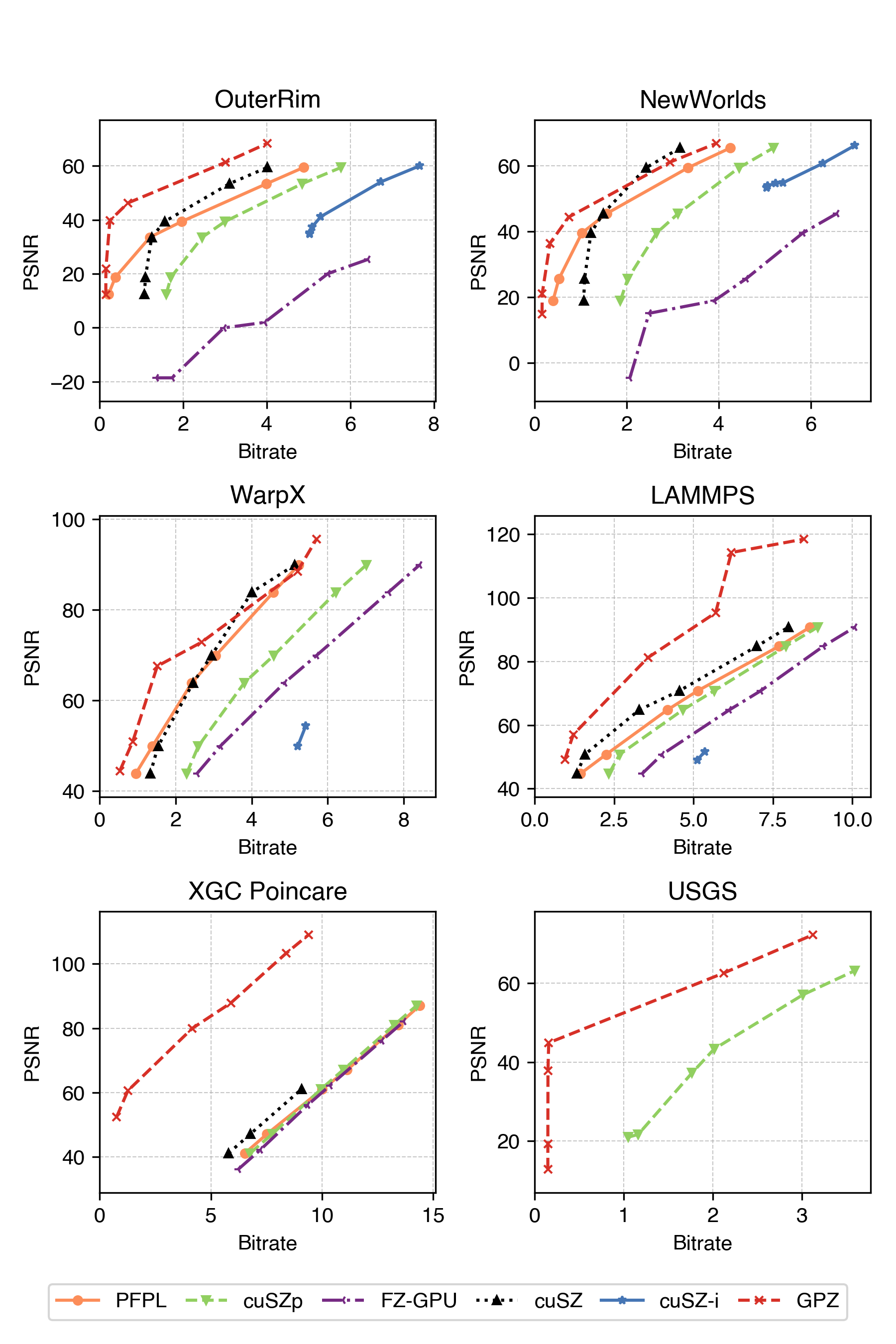}
    \vspace{-2mm}
    \caption{Rate-distortion comparison on six scientific datasets. Higher and more leftward curves represent better performance. GPZ consistently achieves a superior trade-off across all tested data.}
    \label{fig:bit_rate_to_psnr}
    \vspace{-2mm}
\end{figure}

\begin{figure*}[ht!]
    \centering
    \includegraphics[width=\textwidth]{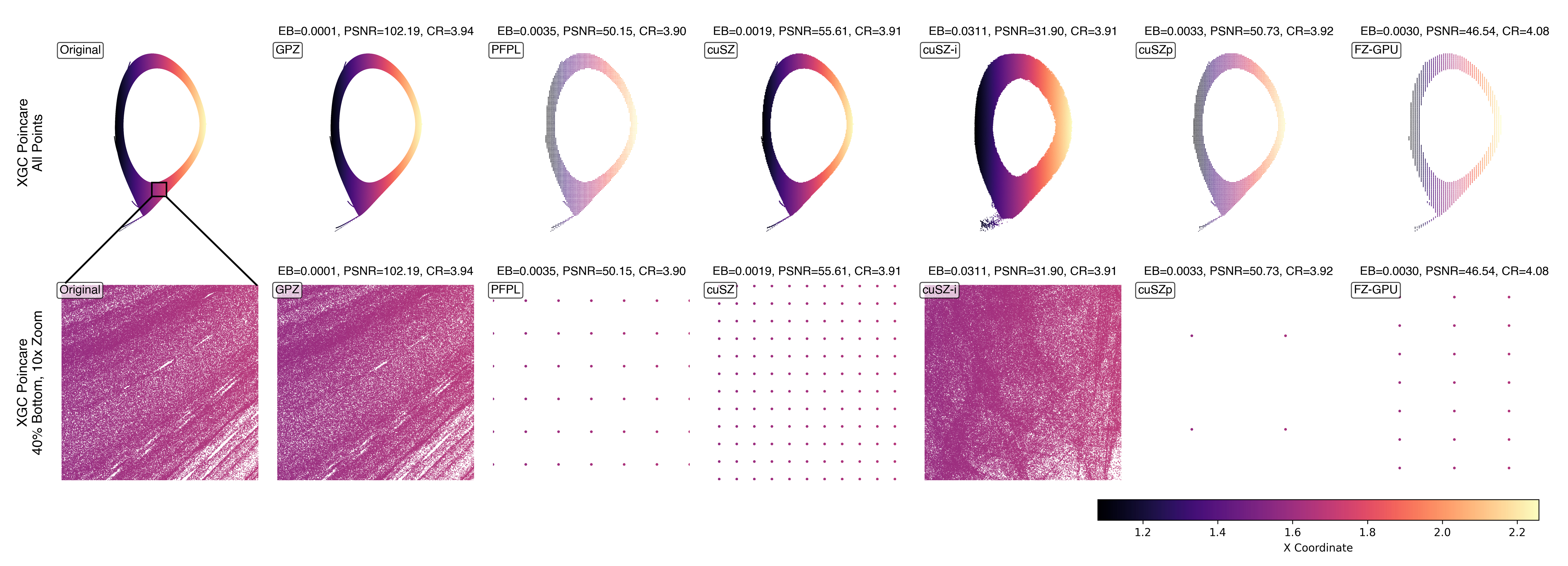}
    \caption{Visual comparison of decompressed XGC Poincare data, with bottom row shows a 10x zoom of the highlighted region. GPZ preserves the fine-grained particle structures, while other methods introduce severe artifacts and data loss.}
    \label{fig:xgc_visual_comparison}
\end{figure*}

As shown in Figure~\ref{fig:bit_rate_to_psnr}, GPZ consistently delivers superior or highly competitive rate-distortion across all six scientific datasets. 
Compressors like FZ-GPU and cuSZ generally fall into the lower-right quadrant confirming the limitation of applying Lorenzo prediction directly on particle datasets.

Numerical metrics like PSNR may not fully capture the preservation of critical scientific features. We therefore visually inspected the decompressed data, using the XGC Poincare dataset shown in Figure~\ref{fig:xgc_visual_comparison}. 
The data decompressed by \textbf{GPZ} is visually almost indistinguishable from the original data, both in the overall Poincaré plot structure and in the 10x magnified view of the particle distribution. In sharp contrast, other compressors introduce severe artifacts which impact post-hoc analysis and visualization tasks.

\section{Conclusion and Future Work}
\label{sec:conclusion}

In this paper, we addressed the critical challenge posed by the massive growth of particle data in modern scientific and commercial applications. We presented GPZ, a novel, high-performance GPU-based lossy compressor designed to bridge the gap between the high throughput of generic GPU compressors and the high compression ratios of particle-specific algorithms. By co-designing a four-stage parallel pipeline with a suite of targeted optimizations for computation, memory, and occupancy, GPZ effectively balances algorithmic efficiency with the architectural constraints of modern GPUs. Our extensive experimental evaluation across diverse datasets and multiple hardware platforms demonstrates that GPZ consistently and substantially outperforms state-of-the-art competitors in end-to-end throughput, compression ratio, and data quality, making it a robust and effective solution for managing large-scale particle data. Future work will focus on integrating GPZ into major simulation frameworks for in-situ data reduction.

\section{Acknowledgments}
\footnotesize
This research was supported by the National Science Foundation under Grant OAC-2104023, OAC-2311875, OAC-2344717, and by the U.S. Department of Energy, Office of Science, Advanced Scientific Computing Research (ASCR), under contract DE-AC02-06CH11357.
This work used the Purdue Anvil cluster through allocation CIS230308 and CIS240192 from the Advanced Cyberinfrastructure Coordination Ecosystem: Services \& Support (ACCESS) program.
This work used ChatGPT and Gemini for language rectification.

\bibliographystyle{ACM-Reference-Format}
\bibliography{citations.bib}


\begin{thebibliography}{29}


\ifx \showCODEN    \undefined \def \showCODEN     #1{\unskip}     \fi
\ifx \showISBNx    \undefined \def \showISBNx     #1{\unskip}     \fi
\ifx \showISBNxiii \undefined \def \showISBNxiii  #1{\unskip}     \fi
\ifx \showISSN     \undefined \def \showISSN      #1{\unskip}     \fi
\ifx \showLCCN     \undefined \def \showLCCN      #1{\unskip}     \fi
\ifx \shownote     \undefined \def \shownote      #1{#1}          \fi
\ifx \showarticletitle \undefined \def \showarticletitle #1{#1}   \fi
\ifx \showURL      \undefined \def \showURL       {\relax}        \fi
\providecommand\bibfield[2]{#2}
\providecommand\bibinfo[2]{#2}
\providecommand\natexlab[1]{#1}
\providecommand\showeprint[2][]{arXiv:#2}

\bibitem[{cuSZ: A GPU-Based Error-Bounded Lossy Compressor for Scientific Data}(2020)]%
        {cusz-git}
\bibfield{author}{\bibinfo{person}{{cuSZ: A GPU-Based Error-Bounded Lossy Compressor for Scientific Data}}.} \bibinfo{year}{2020}\natexlab{}.
\newblock \bibinfo{howpublished}{\url{https://github.com/szcompressor/cuSZ}}.
\newblock
\newblock
\shownote{Online}.


\bibitem[{cuSZ-i: High-Ratio Scientific Lossy Compression on GPUs with Optimized Multi-Level Interpolation}(2024)]%
        {cuszi-git}
\bibfield{author}{\bibinfo{person}{{cuSZ-i: High-Ratio Scientific Lossy Compression on GPUs with Optimized Multi-Level Interpolation}}.} \bibinfo{year}{2024}\natexlab{}.
\newblock \bibinfo{howpublished}{\url{https://github.com/JLiu-1/cusz-I}}.
\newblock
\newblock
\shownote{Online}.


\bibitem[{cuSZp2: A GPU Lossy Compressor with Extreme Throughput and Optimized Compression Ratio}(2024)]%
        {cuszp2-git}
\bibfield{author}{\bibinfo{person}{{cuSZp2: A GPU Lossy Compressor with Extreme Throughput and Optimized Compression Ratio}}.} \bibinfo{year}{2024}\natexlab{}.
\newblock \bibinfo{howpublished}{\url{https://github.com/szcompressor/cuSZp}}.
\newblock
\newblock
\shownote{Online}.


\bibitem[Fallin et~al\mbox{.}(2025)]%
        {pfpl}
\bibfield{author}{\bibinfo{person}{Alex Fallin}, \bibinfo{person}{Noushin Azami}, \bibinfo{person}{Sheng Di}, \bibinfo{person}{Franck Cappello}, {and} \bibinfo{person}{Martin Burtscher}.} \bibinfo{year}{2025}\natexlab{}.
\newblock \showarticletitle{Fast and Effective Lossy Compression on GPUs and CPUs with Guaranteed Error Bounds}. In \bibinfo{booktitle}{\emph{Proceedings of the 39th IEEE International Parallel and Distributed Processing Symposium (IPDPS)}}.
\newblock


\bibitem[{FZ-GPU: A Fast and High-Ratio Lossy Compressor for Scientific Data on GPUs}(2023)]%
        {fzgpu-git}
\bibfield{author}{\bibinfo{person}{{FZ-GPU: A Fast and High-Ratio Lossy Compressor for Scientific Data on GPUs}}.} \bibinfo{year}{2023}\natexlab{}.
\newblock \bibinfo{howpublished}{\url{https://github.com/szcompressor/FZ-GPU}}.
\newblock
\newblock
\shownote{Online}.


\bibitem[{Google Draco}(2024)]%
        {draco}
\bibfield{author}{\bibinfo{person}{{Google Draco}}.} \bibinfo{year}{2024}\natexlab{}.
\newblock \bibinfo{howpublished}{\url{https://google.github.io/draco}}.
\newblock
\newblock
\shownote{Online}.


\bibitem[{GROMACS XTC Format}(2024)]%
        {xtc}
\bibfield{author}{\bibinfo{person}{{GROMACS XTC Format}}.} \bibinfo{year}{2024}\natexlab{}.
\newblock \bibinfo{howpublished}{\url{https://manual.gromacs.org/archive/5.0.4/online/xtc.html}}.
\newblock
\newblock
\shownote{Online}.


\bibitem[Han and Wang(2023)]%
        {coordnet}
\bibfield{author}{\bibinfo{person}{Jun Han} {and} \bibinfo{person}{Chaoli Wang}.} \bibinfo{year}{2023}\natexlab{}.
\newblock \showarticletitle{CoordNet: Data Generation and Visualization Generation for Time-Varying Volumes via a Coordinate-Based Neural Network}.
\newblock \bibinfo{journal}{\emph{IEEE Transactions on Visualization and Computer Graphics}} \bibinfo{volume}{29}, \bibinfo{number}{12} (\bibinfo{year}{2023}), \bibinfo{pages}{4951--4963}.
\newblock
\href{https://doi.org/10.1109/TVCG.2022.3197203}{doi:\nolinkurl{10.1109/TVCG.2022.3197203}}


\bibitem[Heitmann et~al\mbox{.}(2024)]%
        {hacc-new-world}
\bibfield{author}{\bibinfo{person}{Katrin Heitmann}, \bibinfo{person}{Thomas Uram}, \bibinfo{person}{Nicholas Frontiere}, \bibinfo{person}{Salman Habib}, \bibinfo{person}{Adrian Pope}, \bibinfo{person}{Silvio Rizzi}, {and} \bibinfo{person}{Joe Insley}.} \bibinfo{year}{2024}\natexlab{}.
\newblock \showarticletitle{The New Worlds Simulations: Large-scale Simulations across Three Cosmologies}.
\newblock  (\bibinfo{year}{2024}).
\newblock
\href{https://doi.org/10.48550/ARXIV.2406.07276}{doi:\nolinkurl{10.48550/ARXIV.2406.07276}}


\bibitem[Huang et~al\mbox{.}(2024)]%
        {cuszp2}
\bibfield{author}{\bibinfo{person}{Yafan Huang}, \bibinfo{person}{Sheng Di}, \bibinfo{person}{Guanpeng Li}, {and} \bibinfo{person}{Franck Cappello}.} \bibinfo{year}{2024}\natexlab{}.
\newblock \showarticletitle{cuSZp2: A GPU Lossy Compressor with Extreme Throughput and Optimized Compression Ratio} \emph{(\bibinfo{series}{SC '24})}. \bibinfo{publisher}{IEEE Press}, Article \bibinfo{articleno}{15}, \bibinfo{numpages}{18}~pages.
\newblock
\showISBNx{9798350352917}
\href{https://doi.org/10.1109/SC41406.2024.00021}{doi:\nolinkurl{10.1109/SC41406.2024.00021}}


\bibitem[Huang et~al\mbox{.}(2023)]%
        {cuszp}
\bibfield{author}{\bibinfo{person}{Yafan Huang}, \bibinfo{person}{Sheng Di}, \bibinfo{person}{Xiaodong Yu}, \bibinfo{person}{Guanpeng Li}, {and} \bibinfo{person}{Franck Cappello}.} \bibinfo{year}{2023}\natexlab{}.
\newblock \showarticletitle{cuSZp: An Ultra-fast GPU Error-bounded Lossy Compression Framework with Optimized End-to-End Performance}. In \bibinfo{booktitle}{\emph{Proceedings of the International Conference for High Performance Computing, Networking, Storage and Analysis}} \emph{(\bibinfo{series}{SC '23})}. \bibinfo{publisher}{Association for Computing Machinery}, \bibinfo{address}{New York, NY, USA}, Article \bibinfo{articleno}{43}, \bibinfo{numpages}{13}~pages.
\newblock
\showISBNx{9798400701092}
\href{https://doi.org/10.1145/3581784.3607048}{doi:\nolinkurl{10.1145/3581784.3607048}}


\bibitem[Isenburg(2013)]%
        {laszip}
\bibfield{author}{\bibinfo{person}{Martin Isenburg}.} \bibinfo{year}{2013}\natexlab{}.
\newblock \showarticletitle{LASzip: lossless compression of LiDAR data}.
\newblock \bibinfo{journal}{\emph{Photogrammetric Engineering \& Remote Sensing}}  \bibinfo{volume}{79} (\bibinfo{date}{02} \bibinfo{year}{2013}).
\newblock
\href{https://doi.org/10.14358/PERS.79.2.209}{doi:\nolinkurl{10.14358/PERS.79.2.209}}


\bibitem[Li et~al\mbox{.}(2023)]%
        {sperr}
\bibfield{author}{\bibinfo{person}{Shaomeng Li}, \bibinfo{person}{Peter Lindstrom}, {and} \bibinfo{person}{John Clyne}.} \bibinfo{year}{2023}\natexlab{}.
\newblock \showarticletitle{Lossy Scientific Data Compression With SPERR}. In \bibinfo{booktitle}{\emph{2023 IEEE International Parallel and Distributed Processing Symposium (IPDPS)}}. \bibinfo{pages}{1007--1017}.
\newblock
\href{https://doi.org/10.1109/IPDPS54959.2023.00104}{doi:\nolinkurl{10.1109/IPDPS54959.2023.00104}}


\bibitem[Liang et~al\mbox{.}(2022)]%
        {mgardx}
\bibfield{author}{\bibinfo{person}{Xin Liang}, \bibinfo{person}{Ben Whitney}, \bibinfo{person}{Jieyang Chen}, \bibinfo{person}{Lipeng Wan}, \bibinfo{person}{Qing Liu}, \bibinfo{person}{Dingwen Tao}, \bibinfo{person}{James Kress}, \bibinfo{person}{David Pugmire}, \bibinfo{person}{Matthew Wolf}, \bibinfo{person}{Norbert Podhorszki}, {and} \bibinfo{person}{Scott Klasky}.} \bibinfo{year}{2022}\natexlab{}.
\newblock \showarticletitle{MGARD+: Optimizing Multilevel Methods for Error-Bounded Scientific Data Reduction}.
\newblock \bibinfo{journal}{\emph{IEEE Trans. Comput.}} \bibinfo{volume}{71}, \bibinfo{number}{7} (\bibinfo{year}{2022}), \bibinfo{pages}{1522--1536}.
\newblock
\href{https://doi.org/10.1109/TC.2021.3092201}{doi:\nolinkurl{10.1109/TC.2021.3092201}}


\bibitem[Lindstrom(2014)]%
        {zfp}
\bibfield{author}{\bibinfo{person}{Peter Lindstrom}.} \bibinfo{year}{2014}\natexlab{}.
\newblock \showarticletitle{Fixed-rate compressed floating-point arrays}.
\newblock \bibinfo{journal}{\emph{IEEE Transactions on Visualization and Computer Graphics}} \bibinfo{volume}{20}, \bibinfo{number}{12} (\bibinfo{year}{2014}), \bibinfo{pages}{2674--2683}.
\newblock


\bibitem[{MPEG G-PCC}(2024)]%
        {tmc13}
\bibfield{author}{\bibinfo{person}{{MPEG G-PCC}}.} \bibinfo{year}{2024}\natexlab{}.
\newblock \bibinfo{howpublished}{\url{https://github.com/MPEGGroup/mpeg-pcc-tmc13}}.
\newblock
\newblock
\shownote{Online}.


\bibitem[{MPEG V-PCC}(2024)]%
        {tmc2}
\bibfield{author}{\bibinfo{person}{{MPEG V-PCC}}.} \bibinfo{year}{2024}\natexlab{}.
\newblock \bibinfo{howpublished}{\url{https://github.com/MPEGGroup/mpeg-pcc-tmc2}}.
\newblock
\newblock
\shownote{Online}.


\bibitem[{ORNL Summit supercomputer}(2021)]%
        {summit}
\bibfield{author}{\bibinfo{person}{{ORNL Summit supercomputer}}.} \bibinfo{year}{2021}\natexlab{}.
\newblock \bibinfo{howpublished}{\url{https://www.olcf.ornl.gov/summit/}}.
\newblock
\newblock
\shownote{Online}.


\bibitem[{PFPL: a guaranteed-error bound lossy compressor/decompressor that produces bit-for-bit identical files on CPUs and GPUs}(2025)]%
        {pfpl-git}
\bibfield{author}{\bibinfo{person}{{PFPL: a guaranteed-error bound lossy compressor/decompressor that produces bit-for-bit identical files on CPUs and GPUs}}.} \bibinfo{year}{2025}\natexlab{}.
\newblock \bibinfo{howpublished}{\url{https://github.com/burtscher/PFPL}}.
\newblock
\newblock
\shownote{Online}.


\bibitem[Plimpton(1995)]%
        {lammps}
\bibfield{author}{\bibinfo{person}{Steve Plimpton}.} \bibinfo{year}{1995}\natexlab{}.
\newblock \showarticletitle{Fast Parallel Algorithms for Short-Range Molecular Dynamics}.
\newblock \bibinfo{journal}{\emph{J. Comput. Phys.}} \bibinfo{volume}{117}, \bibinfo{number}{1} (\bibinfo{year}{1995}), \bibinfo{pages}{1--19}.
\newblock
\showISSN{0021-9991}


\bibitem[Ren et~al\mbox{.}(2024)]%
        {ren2024particle}
\bibfield{author}{\bibinfo{person}{Congrong Ren}, \bibinfo{person}{Sheng Di}, \bibinfo{person}{Longtao Zhang}, \bibinfo{person}{Kai Zhao}, {and} \bibinfo{person}{Hanqi Guo}.} \bibinfo{year}{2024}\natexlab{}.
\newblock \bibinfo{title}{An Error-Bounded Lossy Compression Method with Bit-Adaptive Quantization for Particle Data}.
\newblock
\showeprint[arxiv]{2404.02826}~[cs.IT]
\urldef\tempurl%
\url{https://arxiv.org/abs/2404.02826}
\showURL{%
\tempurl}


\bibitem[simulation dataset(2024)]%
        {hacc-outerrim}
\bibfield{author}{\bibinfo{person}{{HACC} Outer~Rim simulation dataset}.} \bibinfo{year}{2024}\natexlab{}.
\newblock \bibinfo{howpublished}{\url{https://cosmology.alcf.anl.gov/outerrim/}}.
\newblock
\newblock
\shownote{Online}.


\bibitem[Tao et~al\mbox{.}(2017)]%
        {sz17}
\bibfield{author}{\bibinfo{person}{Dingwen Tao}, \bibinfo{person}{Sheng Di}, \bibinfo{person}{Zizhong Chen}, {and} \bibinfo{person}{Franck Cappello}.} \bibinfo{year}{2017}\natexlab{}.
\newblock \showarticletitle{Significantly improving lossy compression for scientific data sets based on multidimensional prediction and error-controlled quantization}. In \bibinfo{booktitle}{\emph{2017 IEEE International Parallel and Distributed Processing Symposium}}. IEEE, \bibinfo{pages}{1129--1139}.
\newblock


\bibitem[Tian et~al\mbox{.}(2020)]%
        {cusz}
\bibfield{author}{\bibinfo{person}{Jiannan Tian} {et~al\mbox{.}}} \bibinfo{year}{2020}\natexlab{}.
\newblock \showarticletitle{{CuSZ}: An Efficient GPU-Based Error-Bounded Lossy Compression Framework for Scientific Data}. \bibinfo{howpublished}{\url{https://github.com/szcompressor/cuSZ}}. In \bibinfo{booktitle}{\emph{Proceedings of the ACM International Conference on Parallel Architectures and Compilation Techniques}} (Virtual Event, GA, USA) \emph{(\bibinfo{series}{PACT '20})}. \bibinfo{publisher}{Association for Computing Machinery}, \bibinfo{address}{New York, NY, USA}, \bibinfo{pages}{3–15}.
\newblock
\showISBNx{9781450380751}
\newblock
\shownote{Online}.


\bibitem[{USGS 3D Elevation Program}(2024)]%
        {3dep}
\bibfield{author}{\bibinfo{person}{{USGS 3D Elevation Program}}.} \bibinfo{year}{2024}\natexlab{}.
\newblock \bibinfo{howpublished}{\url{https://www.usgs.gov/3d-elevation-program}}.
\newblock
\newblock
\shownote{Online}.


\bibitem[{WarpX}(2024)]%
        {warpx}
\bibfield{author}{\bibinfo{person}{{WarpX}}.} \bibinfo{year}{2024}\natexlab{}.
\newblock \bibinfo{howpublished}{\url{https://ecp-warpx.github.io//}}.
\newblock
\newblock
\shownote{Online}.


\bibitem[Zhang et~al\mbox{.}(2025)]%
        {lcp}
\bibfield{author}{\bibinfo{person}{Longtao Zhang}, \bibinfo{person}{Ruoyu Li}, \bibinfo{person}{Congrong Ren}, \bibinfo{person}{Sheng Di}, \bibinfo{person}{Jinyang Liu}, \bibinfo{person}{Jiajun Huang}, \bibinfo{person}{Robert Underwood}, \bibinfo{person}{Pascal Grosset}, \bibinfo{person}{Dingwen Tao}, \bibinfo{person}{Xin Liang}, \bibinfo{person}{Hanqi Guo}, \bibinfo{person}{Franck Cappello}, {and} \bibinfo{person}{Kai Zhao}.} \bibinfo{year}{2025}\natexlab{}.
\newblock \showarticletitle{LCP: Enhancing Scientific Data Management with Lossy Compression for Particles}.
\newblock \bibinfo{journal}{\emph{Proc. ACM Manag. Data}} \bibinfo{volume}{3}, \bibinfo{number}{1}, Article \bibinfo{articleno}{50} (\bibinfo{date}{Feb.} \bibinfo{year}{2025}), \bibinfo{numpages}{27}~pages.
\newblock
\href{https://doi.org/10.1145/3709700}{doi:\nolinkurl{10.1145/3709700}}


\bibitem[Zhao et~al\mbox{.}(2021)]%
        {interp}
\bibfield{author}{\bibinfo{person}{Kai Zhao}, \bibinfo{person}{Sheng Di}, \bibinfo{person}{Maxim Dmitriev}, \bibinfo{person}{Thierry-Laurent~D. Tonellot}, \bibinfo{person}{Zizhong Chen}, {and} \bibinfo{person}{Franck Cappello}.} \bibinfo{year}{2021}\natexlab{}.
\newblock \showarticletitle{Optimizing Error-Bounded Lossy Compression for Scientific Data by Dynamic Spline Interpolation}. In \bibinfo{booktitle}{\emph{IEEE 37th International Conference on Data Engineering}}. \bibinfo{pages}{1643--1654}.
\newblock
\href{https://doi.org/10.1109/ICDE51399.2021.00145}{doi:\nolinkurl{10.1109/ICDE51399.2021.00145}}


\bibitem[Zhao et~al\mbox{.}(2022)]%
        {mdz}
\bibfield{author}{\bibinfo{person}{Kai Zhao}, \bibinfo{person}{Sheng Di}, \bibinfo{person}{Danny Perez}, \bibinfo{person}{Xin Liang}, \bibinfo{person}{Zizhong Chen}, {and} \bibinfo{person}{Franck Cappello}.} \bibinfo{year}{2022}\natexlab{}.
\newblock \showarticletitle{MDZ: An Efficient Error-bounded Lossy Compressor for Molecular Dynamics}. In \bibinfo{booktitle}{\emph{2022 IEEE 38th International Conference on Data Engineering (ICDE)}}. \bibinfo{pages}{27--40}.
\newblock
\href{https://doi.org/10.1109/ICDE53745.2022.00007}{doi:\nolinkurl{10.1109/ICDE53745.2022.00007}}


\end{thebibliography}

\end{document}